\documentclass[prd,nofootinbib,preprint,superscriptaddress]{revtex4}

\usepackage{amsmath, amssymb, amsthm, graphicx, epsfig, fancyhdr,epsfig, slashed}

\usepackage{tikzsymbols}
\usepackage{natbib}
\usepackage{float}

\usepackage{tikz,xcolor,hyperref}

\usepackage{bm}
\usepackage{url}

\newcommand{\pd}{\partial}

\newcommand{\be}{\begin{equation}}
\newcommand{\ee}{\end{equation}}
\newcommand{\bea}{\begin{eqnarray}}
\newcommand{\eea}{\end{eqnarray}}

\definecolor{lime}{HTML}{A6CE39}
\DeclareRobustCommand{\orcidicon}{
	\begin{tikzpicture}
	\draw[lime, fill=lime] (0,0) 
	circle [radius=0.2] 
	node[white] {{\fontfamily{qag}\selectfont \tiny ID}};
	\draw[white, fill=white] (-0.0625,0.095) 
	circle [radius=0.007];
	\end{tikzpicture}
	\hspace{-2mm}
}

\foreach \x in {A, ..., Z}{\expandafter\xdef\csname orcid\x\endcsname{\noexpand\href{https://orcid.org/\csname orcidauthor\x\endcsname}
			{\noexpand\orcidicon}}
}

\begin{document}


\title{Quintessence Dark Energy from strongly-coupled Higgs mass gap: \\ \it{Local \& Non-local higher-derivative non-perturbative scenarios}}



\author{Marco Frasca\orcidA{}}
\email{marcofrasca@mclink.it}
\affiliation{Rome, Italy}

\author{Anish Ghoshal\orcidB{}}
\email{anish.ghoshal@fuw.edu.pl}
\affiliation{Institute of Theoretical Physics, Faculty of Physics, University of Warsaw, ul. Pasteura 5, 02-093 Warsaw, Poland}

\author{Alexey S. Koshelev\orcidC{}}
\email{ak@inpcs.net}
\affiliation{
Centro de Matem\'atica e Aplica\c{c}\~oes (CMA-UBI) and Departamento de F\'isica\\
Universidade da Beira Interior, Rua Marqu\^es d’Ávila e Bolama 6200-001 Covilh\~a, Portugal
}



\date{\today}

\begin{abstract}
We entertain the possibility that dark energy arises from the Higgs field of quintessence type in the non-perturbative regimes. For this purpose we utilize a set of exact solutions of Higgs field theory recently devised, in terms of Jacobi elliptical function for a massless quartic scalar field, that satisfy a massive dispersion relation. In certain regions of the parameter space, determined by the quartic coupling value, we show that such solutions have the property to give the correct behavior for the equation of state of the dark energy depending on the initial $\theta$ value of periodicity of the Jacobi elliptical function solution. It is seen that on a time scale determined by the Hubble constant and the strength of the self-interaction of the scalar field, when conformal invariance is restored, the equation of
state for the dark energy becomes manifest. Further we investigate scenarios within standard field theory framework and also extending to higher-derivative theories, namely infinite-derivative theory, motivated from p-adic string field theory, and Lee-Wick theories  and in all cases, we find suitable choices of the parameter space leads to dark energy behavior of the Higgs field, which we compare. 
\end{abstract}


\maketitle

\section{Introduction}

We know that during the early stages, the universe suffered an initial period of exponential growth, which wiped out all inhomogeneities and flattened the spacetime, something well known as the inflation. This phase of accelerated expansion rate of the universe is however not unique in the history as after a long period of deceleration, we find that the Universe began to accelerate again, and that too very recently with respect to the lifetime of the universe. We call the unknown agent the ``dark energy'' (DE) which drives this late time acceleration. The cosmological data from the Cosmic Mircowave Background (CMB) \cite{Aghanim:2018eyx} and data from the supernovae \cite{Riess:1998cb,Perlmutter:1998np} strongly hints to the fact that the Universe is partially filled with this DE \cite{Bamba:2012cp}, which is almost about $70\%$ of the total energy density of the Universe. Finding a non-trivial explanation for the acceleration mechanism is very difficult as gravity theories are very strict regarding the matter content. \textit{For example,} the dynamics of early accelerated expansion cannot be driven by ordinary matter but we need a scalar field rolling over a very flat potential mimicking the energy \textit{equation-of-state} of vacuum as proposed in the early 80's as models of inflationary cosmology \cite{inflation}, and to predict and compare to experimental data \cite{planck} we only need to know its potential. 

Similarly, the late time acceleration that we observe or the dark energy cannot be accounted by any ordinary matter but requiring ideas like modified gravity to the action of fundamental scalar fields, or simple explanations like that by Albert\ Einstein to incorporate in a very tiny as well as unnatural cosmological constant (CC) representing the energy density of order of $\rho_{DE}\sim 10^{-119}M_p^4$, where $M_p \simeq 2.435 \times 10^{18}$ GeV is the reduced Planck mass. Future missions like the EUCLID will be able to say a lot on this issue \cite{euclid}. Alternatives to the CC often involve a dynamical origin of the DE, particularly due to presence of some degree of freedom, arising either from a modification of Einstein gravity or from additional scalar fields \cite{Copeland:2006wr,Rinaldi:2014yta}. However the present day value of the energy density $\rho_{DE}$ invokes \textit{fine-tuning problem} in many ways. \textit{For example,} if one assumes DE is a truly a constant like the CC (or a fluid having the equation of state $w\simeq -1$ throughout the whole evolution of the Universe up till today) one will obtain $\rho_{DE} \lll M^4$, where $M$ is any fundamental scale of known physics, such as $M_p\simeq 10^{18}$ GeV, $M_{EW}\sim 10^2$ GeV or $M_{QCD}\sim 0.3 GeV$ \cite{Carroll:2000fy}, therefore leading to big hierarchy problem issues, between $\rho_{DE}$ and other energy densities in the early Universe. This is often called initial value problem for the DE \cite{steinhardt,Dinverno}. Besides, there is also the issue of the so-called coincidence problem here-in \cite{Zlatev:1998tr,Velten:2014nra}. As we know most of the evolution of the universe took place in radiation-dominated and matter-dominated eras, it is indeed rather a big coincidence that current the energy densities of matter (or dust) and DE are of the same \textit{order-of-magnitude}. Finally, we also encounter the Hubble tension problem, suggesting $\gtrsim5\sigma$ discrepancy between the value of the Hubble parameter measured by late-time Universe observations $(z\lesssim 1)$ compared to early-time universe observations $(z\gg1)$ \cite{Verde:2019ivm}.


\noindent The aim of this paper is to propose an intriguing idea that Higgs field with a non-trivial background solution (other than zero) can account for the dark energy. The same field configuration leads to a slowly varying and arbitrarily tiny effective cosmological constant in the late Universe. This conclusion emerges naturally from the analysis of the dynamics of the full Higgs model.  Just to note in this context that there already exist other non-trivial solutions with non-minimal coupling of the Higgs field to gravity, such as the Higgs monopole with static and spherical symmetry studied in \cite{monop}.

Moreover, besides having a local Higgs theory, an inspiration from string theory often entails that strings are non-local objects and infinite-derivative extensions of weakly-coupled non-local Quantum Field Theory (QFT) as studied in Refs. \cite{Taylor:2003gn,Moffat:1990jj,Evens:1990wf,Tomboulis:1997gg, Moffat:2011an,Tomboulis:2015gfa,Kleppe:1991rv,sft1,sft2,sft3,padic1,padic2,padic3,Frampton-padic,marc,Tseytlin:1995uq,Siegel:2003vt,Calcagni:2013eua,Modesto:2011kw,Modesto:2012ga,Modesto:2015foa,Modesto:2017hzl,Taylor:2003gn} has become a very popular and alternative way to address the divergence and the Higgs and gauge hierarchy problem issues in ordinary QFT and in the Standard Model (SM) of particle physics by generalizing the kinetic energy operators (which is second in order) to an ghost-free infinite series of higher order derivatives that are suppressed due the scale of non-locality ($M$) \cite{Krasnikov:1987yj,Biswas:2014yia,Buoninfante:2018mre} with stable Standard Model (SM) Higgs vacuum \cite{Olive:2016xmw,Ghoshal:2017egr,Ghoshal:2020lfd} as have been studied by some of the authors. It was found out that the $\beta$-functions reach a conformal limit\footnote{For other conformal theory applications to cosmology, see \cite{Barman:2021lot} for freeze-in dark matter and \cite{Ghoshal:2022hyc} for inflation. just for some examples.} resolving the issue of Landau-poles \cite{Ghoshal:2020lfd} (see Refs. \cite{Biswas:2014yia,Buoninfante:2018gce,Ghoshal:2018gpq,Krasnikov:2020kgh} for Large Hadron Collider (LHC) phenomenology, astrophysical implications, dimensional transmutation and dark matter and other phenomenology and proton decays in Grand Unified Theories (GUT) in this higher-derivative framework). For such higher-derivative Higgs theory the non-perturbative strongly-coupled regimes and exact $\beta-$functions, and conditions of confinement, have been actively investigated very recently by some of the authors \cite{Frasca:2020ojd,Frasca:2020jbe,Frasca:2021iip,Frasca:2022duz,Frasca:2022lwp}
\footnote{The unitarity issues are well addressed and understood in Euclidean space and via using Cutkosky rules, the results obtained in Eucliedean can be analytically continued to the Minkowski space, see Refs. \cite{Pius:2016jsl,Briscese:2018oyx,Briscese:2021mob,Koshelev:2021orf}. Therefore, we assume an Euclidean metric everywhere in this work. Models of the kind which we consider in our paper
originate in string field theory. Technically speaking they cannot be avoided and are in the literature since the original Witten's papers on String Field Theory (SFT) (see Refs. \cite{sft1,padic2,padic3}. In Ref. \cite{Koshelev:2021orf} it was proven
that unitarity cannot be maintained unless
you modify the Feynman graph computation prescription as proposed in other references above. Note that a modified prescription restores a standard computation scheme in case of local theories. In short, given the imminent appearance of such models in String Field Theory the model we consider here is physically relevant.}.

In order to study the non-perturbative regimes in Higgs theories, local and non-local, we revert to utilising exact solutions to the Higgs theory, as found in terms of Jacobi elliptical functions, following Ref.\cite{Frasca:2015yva}. In this case, the Green's functions of the theory are analytically expressed, and therefore it is easy to understand the effect on the interaction via the background solution which remains valid even in the strongly coupled regimes. This technique has been recently applied to QCD, in Refs.\cite{Frasca:2021yuu,Frasca:2021zyn,Chaichian:2018cyv} and to the Higgs sector of the Standard Model in Ref.\cite{Frasca:2015wva},  and other classes of quantum field theories over the decades in Refs. \cite{Frasca:2019ysi, Chaichian:2018cyv, Frasca:2017slg, Frasca:2016sky, Frasca:2015yva, Frasca:2015wva, Frasca:2013tma, Frasca:2012ne, Frasca:2009bc, Frasca:2010ce, Frasca:2008tg, Frasca:2009yp, Frasca:2008zp, Frasca:2007uz, Frasca:2006yx, Frasca:2005sx, Frasca:2005mv, Frasca:2005fs}. Recently some of the authors studied this in context to non-perturbative hadronic contribution to muon (g-2)$_{\mu}$ magnetic moment estimation \cite{Frasca:2021yuu}, to prove confinement in QCD \cite{Frasca:2021zyn,Frasca:2022lwp}, non-perturbative false vacuum decay \cite{Frasca:2022kfy,Calcagni:2022gac,Calcagni:2022tls} and to study mass gap and confinement in string-inspired infinite-derivative and Lee-Wick theories \cite{Frasca:2020jbe,Frasca:2020ojd,Frasca:2021iip,Frasca:2022gdz}.
The main technique we will rely on is originally devised by Bender, Milton, Savage and the method for Dyson-Schwinger equations as shown in \cite{Bender:1999ek} that is widely used in the aforementioned studies. 

The paper is organized as follows: in section II we briefly review the exact background solution for the Higgs theory in the strongly-coupled regimes, in terms of Jacobi Elliptical function, in section III, we estimate the dark energy equation equation of state, in section IV, we show both the local and non-local Higgs field acts as Quintessence field type and may behave as dark energy at late times. Finally we end with a detailed discussions of our investigation in section V.

\medskip

\section{Exact solutions of Scalar Field: Review}

\subsection{Local Theory}

The equation for the quintessence field $Q$ has the general form in FRW background
\begin{equation}
\label{eq:Q}
   \ddot Q+3H\dot Q+V'(Q)=0
\end{equation}
where $H$ is the Hubble constant and $V(Q)$ the potential with the prime meaning a derivative with respect to $Q$ and the dot is the derivative with respect to time. For a general scalar field on a flat space-time it is possible that a mass gap develops. Indeed, let us consider the case of a quartic scalar field as
\begin{equation}
   \partial^2\phi+\lambda\phi^3=0
\end{equation}
being $\lambda$ the coupling. An exact set of solutions is yielded by \cite{Frasca:2009bc,Frasca:2013tma}
\begin{equation}
\label{eq:exeq}
   \phi(x)=\mu\left(\frac{2}{\lambda}\right)\mathrm{sn}(p\cdot x+\theta,-1)
\end{equation} 
with $\mu$ and $\theta$ two integration constants and sn a Jacobi elliptic function, provided that the following dispersion relation holds
\begin{equation}
\label{eq:dr}
    p^2=\mu^2\sqrt{\frac{\lambda}{2}}
\end{equation}
that appears to be that of a massive field. The corresponding quantum field theory, with such a field pervading all the space-time, maintains the mass gap and adds a 
Kaluza-Klein spectrum on the free particles. This theory has a trivial infrared fixed point \cite{Frasca:2013tma}. We are going to show that such solutions have the property to behave as a quintessence field with the equation of state proper to the cosmological term in the Einstein equations.

Our technique will be based on the fact that the ratio $H/m_0$, being $m_0=\mu(\lambda/2)^\frac{1}{4}$ is greatly lesser than unity and so, it can be used as a development parameter 
for a pertrubation series. Then, we firstly exploit the behavior of the theory at the leading order.

\subsection{Non-local Background Solution}

We begin with the action for the infinite derivative Higgs as given by \cite{Biswas:2014yia}:
\begin{equation} \label{Action}
    S = \int d^4 x\ \left(-\frac{1}{2} \phi e^{f(\Box)}(\Box+m^2)\phi -\frac{\lambda}{4!}\phi^{4}\right)
\end{equation} 
Here the normalization of $\phi$ field is chosen in such a manner that the residue at the $p^2=m^2$ pole is basically unity.
$\Box= \eta_{\mu\nu}\partial^{\mu}\partial^{\nu}$ $(\mu, \nu=0,1,2,3$) with the convention of the metric signature $(+,-,-,-)$, 
  $m$ is the mass of the scalar particle, 
  and $M$ is the energy scale of the non-locality usually considered to be below the Planck scale. The kinetic energy terms are generalized with series of higher derivatives suppressed by the non-local energy scale $M$, however the scalar self-interaction is the typical quartic potential one. Most importantly, the theory reduces to the standard local field theory in the limit of $M \to \infty$. $e^{f(\Box)}$ is the non local form-factor chosen in such a manner that it does not incorporate any new poles in the propagator,
  usually exponential of trancesdal function does this job however various proposals have been explored a lot in the literature (see Ref. \cite{Edholm:2016hbt}). In most of the cases henceforth in this paper we will work with generalised $f(\Box)$, and then we will show and compare our results for some specific choices.

In Euclidean space ($p^0\rightarrow ip_{E} ^0$) the propagator is:
\begin{equation} \label{Prop}
\Pi(p^2)=-\frac{ie^{f(-p_E^2)}}{p_E^2+m^2}
\end{equation}
and the vertex factor is given by $-i\lambda$. It is easy to appreciate that the exponential suppression of the propagators for $p_E^2 > M^2$ regionin the above equation, leads to UV softening of the quantum corrections at energies higher than $M$ (actually all the $\beta-$functions vanish beyond the scale of non-locality M thereby reaching an asymptotically conformal limit in the UV) as shown explicitly in Ref. \cite{Ghoshal:2017egr,Ghoshal:2020lfd}. 

\subsubsection{Iterative approach}

Usage of iterative technique to solve infinite-derivative PDEs has been put forward in Refs.~\cite{Volovich(2003),Vladimirov:2003kg}. We apply these ideas in the framework of strongly coupled non-local theories. In the non-local case, we have the equation of motion
\begin{equation} \label{e}
    \Box f(\Box)\phi=-\lambda\phi^3
\end{equation}
where $f(\Box)$ is the non-local form factor that is expected to be unity in the limit of the non-local scale running to infinity. 

We can solve this by iterative techniques. In order to do this, we note the following identity
\begin{equation}
    (\operatorname{sn}(x,i))''=-2\operatorname{sn}^3(x,i).
\end{equation}
This implies that we are able to write the following Fourier series 
\begin{eqnarray}
    \phi_0^3(x)&=&\mu^3\left(2/\lambda\right)^\frac{3}{4}\mathrm{sn}^3(p\cdot x+\theta,i) \\
    &=&-\frac{1}{2}\mu^3\left(2/\lambda\right)^\frac{3}{4}(\mathrm{sn}(p\cdot x+\theta,i))'' \nonumber \\
    &=&\mu^3\left(2/\lambda\right)^\frac{3}{4}
    \frac{\pi^3}{4K^3(i)}\sum_{n=0}^\infty(-1)^n(2n+1)^2\frac{e^{-\left(n+\frac{1}{2}\right)\pi}}{1+e^{-(2n+1)\pi}}
    \sin\left((2n+1)\frac{\pi}{2K(i)}(p\cdot x+\theta)\right), \nonumber
\end{eqnarray}
where $K(i)$ is the complete elliptic integral of the first kind.
Therefore, if we use, as a first iterate, eq.(\ref{eq:exeq}) that holds in the local limit, we will have to solve
\begin{equation}
 \Box f(\Box)\phi_1=-\lambda\phi_0^3.   
\end{equation}
This equation can be solved immediately by writing down
\begin{equation}
    \phi_1=\sum_{n=0}^\infty a_n\sin\left((2n+1)\frac{\pi}{2K(i)}(p\cdot x+\theta)\right).
\end{equation}
This will yield
\begin{align}
    a_n=f\left(-(2n+1)^2\frac{\pi^2}{4K^2(i)}p^2\right)\left(-p^2(2n+1)^2\frac{\pi^2}{4K^2(i)}\right)^{-1}
    \mu^3\left(8\lambda\right)^\frac{1}{4} \nonumber \\
    \frac{\pi^3}{4K^3(i)}(-1)^n(2n+1)^2\frac{e^{-\left(n+\frac{1}{2}\right)\pi}}{1+e^{-(2n+1)\pi}}.
\end{align}
Working on-shell with the dispersion relation (\ref{eq:dr}), this becomes
\begin{equation}
    a_n=
    \mu\left(32/\lambda\right)^\frac{1}{4}\frac{\pi}{K(i)}f^{-1}\left(-(2n+1)^2\frac{\pi^2}{4K^2(i)}\mu^2\sqrt{\lambda/2}\right)
    (-1)^n\frac{e^{-\left(n+\frac{1}{2}\right)\pi}}{1+e^{-(2n+1)\pi}}.
\end{equation}
We see immediately that this solution is strongly damped in the limit $\lambda\rightarrow\infty$. Besides, one sees that higher harmonics are exponentially damped both by the non-local factor and the by numerical contributions arsing from the original series for the Jacobi sn function.

We can evaluate the next iterate by observing that the series we obtained can be written as
\begin{equation}
\label{eq:fs}
    \phi_1(x)=\frac{1}{2i}\sum_{n=-\infty}^\infty a_n e^{(2n+1)\frac{i\pi}{2K(i)}(p\cdot x+\theta)}.
\end{equation}
Then, for the square one has
\begin{equation}
    \phi_1^2(x)=\frac{1}{4}\sum_{n=-\infty}^\infty c_n e^{(2n+1)\frac{i\pi}{2K(i)}(p\cdot x+\theta)}
\end{equation}
being
\begin{equation}
    c_k=\sum_{i=-\infty}^\infty a_ia_{k-i}.
\end{equation}
For the cube, it is
\begin{equation}
    \phi_1^3(x)=-\frac{1}{8i}\sum_{n=-\infty}^\infty d_n e^{(2n+1)\frac{i\pi}{2K(i)}(p\cdot x+\theta)}
\end{equation}
being
\begin{equation}
    d_k=\sum_{i=-\infty}^\infty c_ia_{k-i}=\sum_{i=-\infty}^\infty\sum_{j=-\infty}^\infty
    a_ja_{i-j}a_{k-i}.
\end{equation}
We get, rewriting eq.(\ref{eq:fs}) as
\begin{equation}
    \phi_2(x)=\frac{1}{2i}\sum_{n=-\infty}^\infty b_n e^{(2n+1)\frac{i\pi}{2K(i)}(p\cdot x+\theta)},
\end{equation}
the identity
\begin{eqnarray}
  &&\frac{1}{2i}f\left(-(2n+1)^2\frac{\pi^2}{4K^2(i)}\mu^2\sqrt{\lambda/2}\right)\left(-(2n+1)^2\frac{\pi^2}{4K^2(i)}\mu^2\sqrt{\lambda/2}\right)b_n= \nonumber \\
  &&-\frac{1}{8i}\lambda \sum_{i=-\infty}^\infty c_ia_{n-i}= \nonumber \\
  &&-\frac{1}{8i}\lambda\sum_{i=-\infty}^\infty\sum_{j=-\infty}^\infty
    a_ja_{i-j}a_{n-i}.
\end{eqnarray}


This yields
\begin{eqnarray}
    b_n&=&a_n(1+e^{-(2n+1)\pi})\times \\
    &&\sum_{k=-\infty}^\infty\sum_{j=-\infty}^\infty
    f^{-1}\left(-(2j+1)^2\frac{\pi^2}{4K^2(i)}\mu^2\sqrt{\lambda/2}\right)
    f^{-1}\left(-(2(k-j)+1)^2\frac{\pi^2}{4K^2(i)}\mu^2\sqrt{\lambda/2}\right)
    \times \nonumber \\
    &&
    f^{-1}\left(-(2(n-k)+1)^2\frac{\pi^2}{4K^2(i)}\mu^2\sqrt{\lambda/2}\right)
    \frac{1}{1+e^{-(2j+1)\pi}}
    \frac{1}{1+e^{-(2(k-j)+1)\pi}}
    \frac{1}{1+e^{-(2(n-k)+1)\pi}}. \nonumber
\end{eqnarray}

We can conclude that the higher harmonics could be heavily damped with respect to the leading one, for a proper choice of the form factor $f$, due to their position in the spectrum and the coupling running to infinity. We exploit this fact in the next section.

\subsubsection{Approximate solution}

The above iterative solution suggests that the higher order excitation are heavily damped by the non-local factor, if properly chosen, both for being higher harmonics and for the coupling running to infinity. Indeed, let us consider the following approximation
\begin{eqnarray}
\label{eq:app}
    \phi(x)&=&c_0
    \sin\left(\frac{\pi}{2K(i)}(p\cdot x+\theta)\right)+
    \nonumber \\
    &&c_1
    \sin\left(\frac{3\pi}{2K(i)}(p\cdot x+\theta)\right)+\ldots.
\end{eqnarray}
On the lhs of the equation of motion we retain both terms while we stop at the first one for the rhs, this gives
\begin{eqnarray}
    &&-p^2\frac{\pi^2}{4K^2(i)}c_0\sin\left(\frac{\pi}{2K(i)}(p\cdot x+\theta)\right)
    f\left(-\frac{\pi^2}{4K^2(i)}p^2\right)\nonumber \\
    &&-p^2\frac{9\pi^2}{4K^2(i)}c_1\sin\left(\frac{3\pi}{2K(i)}(p\cdot x+\theta)\right)f\left(-\frac{9\pi^2}{4K^2(i)}p^2\right)+\ldots= \nonumber \\
    &&-\lambda c_0^3
    \left[\frac{3}{4}\sin\left(\frac{\pi}{2K(i)}(p\cdot x+\theta)\right)-\frac{1}{4}
    \sin\left(\frac{3\pi}{2K(i)}(p\cdot x+\theta)\right)
    \right]+\ldots
\end{eqnarray}
From this, we will get
\begin{eqnarray}
   &&p^2\frac{\pi^2}{4K^2(i)}f\left(-\frac{\pi^2}{4K^2(i)}p^2\right)=\frac{3}{4}\lambda c_0^2 \nonumber \\
   &&c_1=-\frac{\lambda}{4}c_0^3\left(p^2\frac{9\pi^2}{4K^2(i)}\right)^{-1}
   f\left(-\frac{9\pi^2}{4K^2(i)}p^2\right)^{-1}.
\end{eqnarray}
We can fix $c_0$ in such a way that, for the non-local mass scale running to infinity, we recover the local dispersion relation. This will give
\begin{equation}
    c_0^2=\frac{4}{3}\mu^2\frac{\pi^2}{4K^2(i)}\frac{1}{\sqrt{2\lambda}}.
\end{equation}
Therefore, we are able to estimate $c_1$. We will get
\begin{equation}
    c_1=-
    \frac{\pi}{54K(i)}\frac{\mu}{(2\lambda)^\frac{1}{4}}
    f\left(-\frac{\pi^2}{4K^2(i)}p^2\right)f\left(-\frac{9\pi^2}{4K^2(i)}p^2\right)^{-1}.
\end{equation}
In order to complete this estimation, we evaluate $p^2$ to its local value yielding
\begin{equation}
    c_1=-
    \frac{\pi}{54K(i)}\frac{\mu}{(2\lambda)^\frac{1}{4}}
    f\left(-\frac{\pi^2}{4K^2(i)}\mu^2\sqrt{\lambda/2}\right)f\left(-\frac{9\pi^2}{4K^2(i)}\mu^2\sqrt{\lambda/2}\right)^{-1}.
\end{equation}
As expected, the second harmonic is heavily damped and the leading harmonic dominates the solution. We just point out that the dispersion relation takes the form
\begin{equation}
\label{eq:disp}
   p^2f\left(-\frac{\pi^2}{4K^2(i)}p^2\right)=\mu^2\sqrt{\frac{\lambda}{2}}.
\end{equation}
This generalizes the result of the local theory to the non-local infinite-derivative case.

\subsubsection{Green function}

In the limit of the coupling going to infinity, the next-to-leading order equation takes the form
\begin{equation}
\Box f(\Box)\phi_1+3\lambda\phi_0^2(x)\phi_1=j    
\end{equation} 
where $j$ is some external current that we can expand in a Fourier series. We will have
\begin{eqnarray}
    \phi_0^2(x)&=&\left[c_0
    \sin\left(\frac{\pi}{2K(i)}(p\cdot x+\theta)\right)+\right.
    \nonumber \\
    &&\left.c_1
    \sin\left(\frac{3\pi}{2K(i)}(p\cdot x+\theta)\right)+\ldots.\right]^2= \nonumber \\
    &&c_0^2\sin^2\left(\frac{\pi}{2K(i)}(p\cdot x+\theta)\right)+ \nonumber \\
    &&2c_0c_1\sin\left(\frac{\pi}{2K(i)}(p\cdot x+\theta)\right)\sin\left(\frac{3\pi}{2K(i)}(p\cdot x+\theta)\right)+ \nonumber \\
    &&c_1^2\sin^2\left(\frac{3\pi}{2K(i)}(p\cdot x+\theta)\right)+\ldots= \nonumber \\
    &&\frac{c_0^2+c_1^2}{2}-\frac{1}{2}c_0^2\cos\left(\frac{\pi}{K(i)}(p\cdot x+\theta)\right) \nonumber \\
    &&-\frac{c_1^2}{2}\cos\left(\frac{3\pi}{K(i)}(p\cdot x+\theta)\right)+\nonumber \\
    &&c_0c_1\left(\sin\left(\frac{2\pi}{K(i)}(p\cdot x+\theta)\right)
    +\sin\left(\frac{\pi}{K(i)}(p\cdot x+\theta)\right)
    \right)+\ldots.
\end{eqnarray}
So, we see that
\begin{equation}
    \Box f(\Box)\phi_1+\frac{3}{2}\lambda(c_0^2+c_1^2)\phi_1
    +\ldots=j(x),
\end{equation}
the theory develops a mass gap and higher excited states are strongly damped in the $\lambda\rightarrow\infty$ limit.

\medskip

\section{Dark energy equation of state}

Firstly, we note that, using eq.(\ref{eq:exeq0}) and the dispersion relation given in eq.(\ref{eq:dr}), $\rho \sim \mu^4/2$, therefore, $\mu$ could be fixed to $\mu \sim O(10^{-1})$ by the value of the dark energy density, $\rho_{DE} \sim meV$. We do not go into details on such normalization issue instead we like to study the evolution of the scalar field such that if it may mimic dark energy equation-of-state.

\subsection{Local case}

Once such exact solutions are given, we can evaluate the equation of state for dark energy. For this aim let us just consider the rest frame with ${\bm p}=0$, then 
from Eq.(\ref{eq:exeq}) one has
\begin{equation}
\label{eq:exeq0}
   \phi(t,0)=\mu\left(\frac{2}{\lambda}\right)^\frac{1}{4}\mathrm{sn}(m_0 t+\theta,-1)
\end{equation}
We just note that the pressure is
\begin{equation}
   p=\frac{1}{2}\dot\phi^2-\frac{\lambda}{4}\phi^4
\end{equation}
and the density is
\begin{equation}
   \rho=\frac{1}{2}\dot\phi^2+\frac{\lambda}{4}\phi^4.
\end{equation}
So,
\begin{equation}
  w=\frac{p}{\rho}=\frac{\frac{1}{2}\dot\phi^2-\frac{\lambda}{4}\phi^4}{\frac{1}{2}\dot\phi^2+\frac{\lambda}{4}\phi^4}.
\end{equation}
This yields
\begin{equation}
  w=\frac{p}{\rho}=1-2\mathrm{sn}^4(m_0 t+\theta,-1).
\end{equation}
This result shows that, if the mass gap goes to zero and conformal invariance is restored, we get $w=1-2\mathrm{sn}^4(\theta,-1)$. So, depending on the phase $\theta$, we get back the equation of state for the dark energy.
\begin{figure}[H]
\centering
\includegraphics[width=0.65\textwidth]{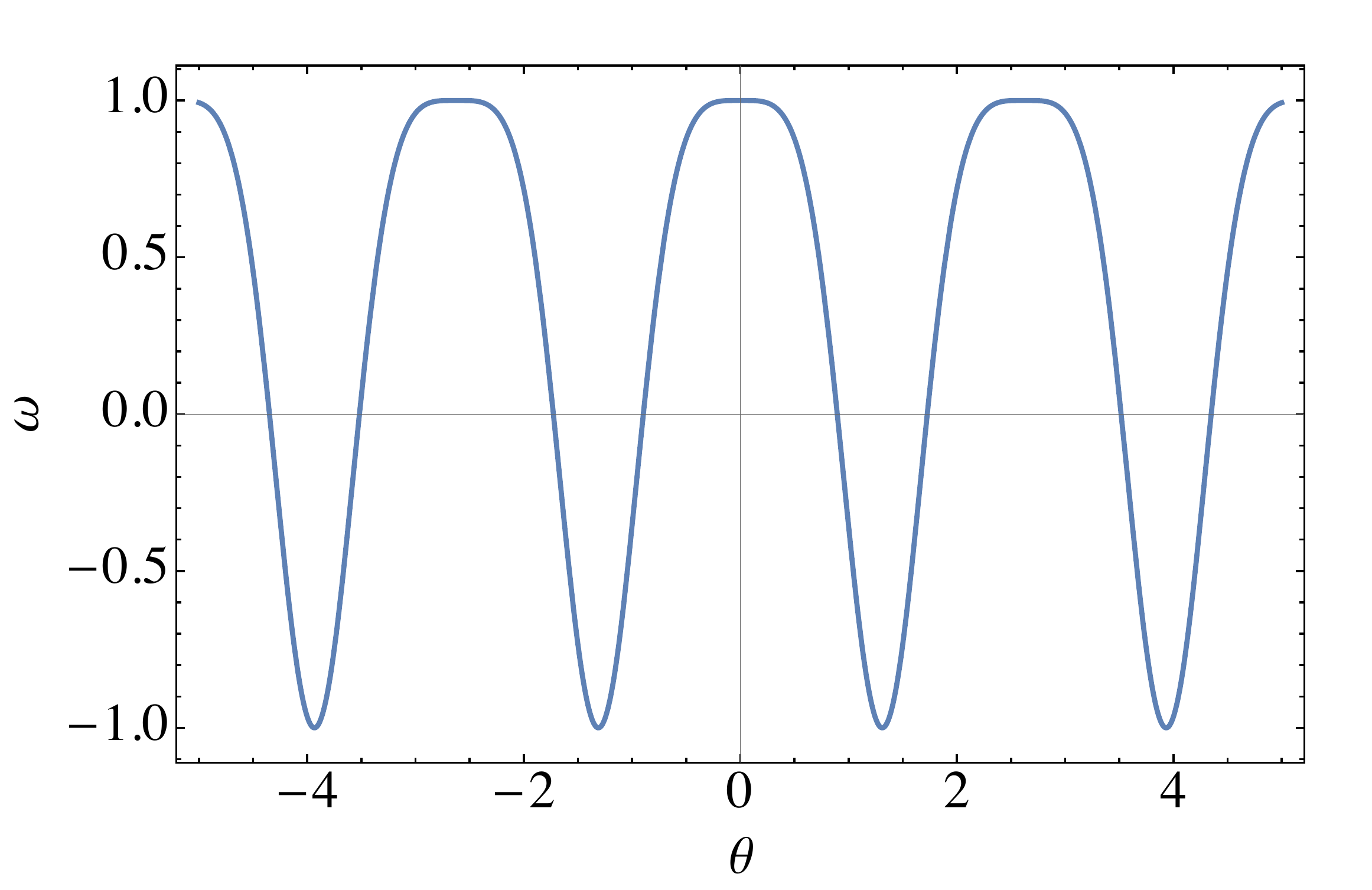}
\caption{\it Variation of Higgs field equation of state $\omega$ with periodicity $\theta$ for the local case.\label{fig1}}
\end{figure} 
We see that complete dark energy corresponds to the values $\theta=\pm(2n+1)K(i)$ where $n\in\mathbb{Z}$. We assume for the time variations that is on a sufficiently small scale with respect to the mass gap $m_0t\approx 0$.

\subsection{Non-local case}

The general equations for pressure $p$ and energy density $\rho$ in the non-local have been derived in Refs. \cite{Koshelev:2010bf,Koshelev:2020fok}. The energy density and the pressure for a perfect fluid in a Friedman-Robertson-Walker metric are \cite{Koshelev:2010bf}
\begin{equation}
\begin{array}{rcl}
\displaystyle \varrho&=&\displaystyle\frac1{2}\sum_{n=1}^\infty
f_n\sum_{l=0}^{n-1} \Bigl(\pd_t\Box^l \phi \pd_t\Box^{n-1-l} \phi +
\Box^l \phi \Box^{n-l} \phi \Bigr)-{}\\[2.7mm]&&\displaystyle {}-
\frac12\phi f(\Box)\phi+V_{int}(\phi),\\[2.7mm]
\displaystyle p&=&\displaystyle\frac1{2}\sum_{n=1}^\infty
f_n\sum_{l=0}^{n-1}\left(\pd_t\Box^l \phi \pd_t\Box^{n-1-l} \phi -
\Box^l \phi \Box^{n-l} \phi
\right)+{}\\[2.7mm]&&\displaystyle{}+\frac12 \phi f(\Box)\phi-V_{int}(\phi),
\end{array}\label{ep_sol}
\end{equation}
where we have taken for the non-local form factor $f(\square)=\Sigma_{k\geq0} f_k\Box^k$ with $f_k\in\mathbb{R}$ and $\Box$ is the Beltrami operator depending on metric. Looking for the first correction terms coming from the higher derivative modifications to local theory, we will get:
\begin{equation}
\rho=\frac{1}{2}\dot{\phi}+V+\frac{1}{M^2}\left(\dot{\phi}\partial_t(\partial_t^2+3H\partial_t)\phi+((\partial_t^2+3H\partial_t)\phi)^2\right)+\ldots\,,
\end{equation}
and, in our approximation, we consider the corrections arising from the metric to become negligible small, in the limit of large non-local mass scale $M$ with respect to the Hubble constant $H$. A similar argument applies for the expression for pressure as well. This particular choice runs similarly to the one given in the analysis for non-local gauge theories in Refs.~\cite{Biswas:2014yia,Ghoshal:2017egr,Frasca:2021iip,Ghoshal:2020lfd,Frasca:2020ojd} which is the essence of considering the effects of non-locality as corrections to local physics including that in collider studies for a simplified study. A detailed discussion is beyond the scope of the present paper.
%

Therefore, given the approximate solution in eq.(\ref{eq:app}) limited just to the first term, 
for the non-local case we just note that the pressure is
\begin{equation}
   p=\frac{1}{2}f(\Box)\dot\phi^2-\frac{\lambda}{4}\phi^4
\end{equation}
and the energy density is
\begin{equation}
   \rho=\frac{1}{2}f(\Box)\dot\phi^2+\frac{\lambda}{4}\phi^4.
\end{equation}
In the rest frame we take $\Box\rightarrow\partial^2_t$. So, the equation-of-state
\begin{equation}
  w=\frac{p}{\rho}=\frac{\frac{1}{2}f(\partial^2_t)\dot\phi^2-\frac{\lambda}{4}\phi^4}{\frac{1}{2}f(\partial^2_t)\dot\phi^2+\frac{\lambda}{4}\phi^4}.
\end{equation}
This yields, in the strong coupling limit,
\begin{equation}
    w=\frac{\frac{1}{2}f\left(-\frac{\pi^2}{4K^2(i)}p_0^2\right)c_0^2\frac{\pi^2}{4K^2(i)}p_0^2 \cos^2\left(\frac{\pi}{2K(i)}(p_0 t+\theta)\right)-\frac{\lambda}{4}c_0^4\sin^4\left(\frac{\pi}{2K(i)}(p_0 t+\theta)\right)}{\frac{1}{2}f\left(-\frac{\pi^2}{4K^2(i)}p_0^2\right)c_0^2\frac{\pi^2}{4K^2(i)}p_0^2\cos^2\left(\frac{\pi}{2K(i)}(p_0 t+\theta)\right)+\frac{\lambda}{4}c_0^4\sin^4\left(\frac{\pi}{2K(i)}(p_0 t+\theta)\right)}.
\end{equation}
This yields
\begin{equation}
    w=\frac{\frac{1}{2}\mu^2\sqrt{\frac{\lambda}{2}}c_0^2\frac{\pi^2}{4K^2(i)} \cos^2\left(\frac{\pi}{2K(i)}(p_0 t+\theta)\right)-\frac{\lambda}{4}c_0^4\sin^4\left(\frac{\pi}{2K(i)}(p_0 t+\theta)\right)}{\frac{1}{2}\mu^2\sqrt{\frac{\lambda}{2}}c_0^2\frac{\pi^2}{4K^2(i)}\cos^2\left(\frac{\pi}{2K(i)}(p_0 t+\theta)\right)+\frac{\lambda}{4}c_0^4\sin^4\left(\frac{\pi}{2K(i)}(p_0 t+\theta)\right)}.
\end{equation}
Finally,
\begin{equation}
\label{eq:wnl}
    w=\frac{
    \cos^2\left(\frac{\pi}{2K(i)}(p_0 t+\theta)\right)-\frac{2}{3}\sin^4\left(\frac{\pi}{2K(i)}(p_0 t+\theta)\right)
    }
    {
    \cos^2\left(\frac{\pi}{2K(i)}(p_0 t+\theta)\right)+\frac{2}{3}\sin^4\left(\frac{\pi}{2K(i)}(p_0 t+\theta)\right)
    }.
\end{equation}

\begin{figure}[H]
\centering
\includegraphics[width=0.65\textwidth]{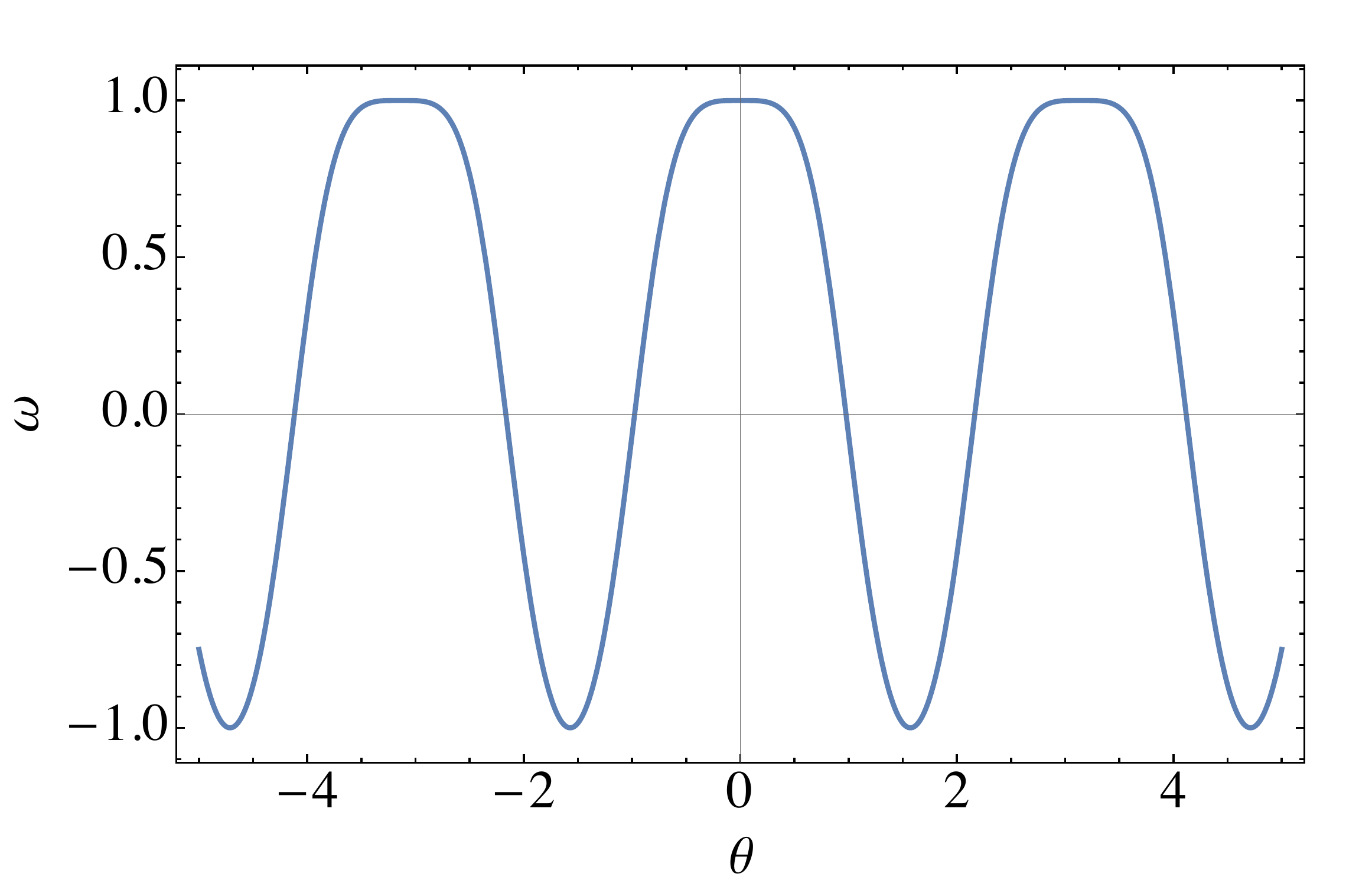}
\caption{\it Variation of Higgs field equation of state $\omega$ with periodicity $\theta$ for the non-local case.\label{fig2}}
\end{figure}

Complete dark energy era corresponds to the values $\theta=\pm\pi/2+2n\pi$ being $n\in\mathbb{Z}$. Again, we are assuming that the time variations can be omitted on sufficiently small scales with respect to the mass gap $p_0t\approx 0$. Here $p_0$ is obtained by solving eq.(\ref{eq:disp}).

The dispersion relation (\ref{eq:disp}) is rather general. It could apply to the Lee-Wick theory as well \cite{LeeWick:1969,Grinstein:2007mp,Cutkosky:1969fq,Anselmi:2017yux,Grinstein:2008bg,Carone:2008bs,Carone:2008iw,LWpheno,Modesto:2016ofr}. When the the non-local factor $f(\Box)=e^{-\frac{\Box}{M^2}}$ with $M$ the non-local scale mass is truncated it at the first order in the exponential series giving $1-\Box/M^2$ we reach the Lee-Wick limit of the theory. This will give rise to the dispersion relation
\begin{equation}
\label{eq:dispLW}
   p^2\left(1+\frac{\pi^2}{4K^2(i)}p^2\right)=\mu^2\sqrt{\frac{\lambda}{2}}.
\end{equation}
The result given in eq.(\ref{eq:wnl}) is left untouched by this particular choice of the non-local factor as we have obtained it for the general case. In turn, this implies that the dark energy is a generic effect of this kind of non-local theories.

 

\medskip

\section{Quintessence equation}

In order to understand how the different physical scales, $m_0$ and the Hubble constant $H$, enter into our discussion, we get explicit solutions to eq.(\ref{eq:Q}). This grants the consistency of our analysis.  

\subsection{Constant solution}

We consider the equation
\begin{equation}
    \ddot\phi+3H\dot\phi+\lambda\phi^3=0,
\end{equation}
and write the solution in the form
\begin{equation}
    \phi(t)=e^{-\frac{3}{2}Ht}q(t).
\end{equation}
We will get
\begin{equation}
\label{eq:q(t)}
    \ddot q-\frac{9}{4}H^2q+\lambda e^{-3Ht}q^3=0.
\end{equation}
This equation can be rewritten as
\begin{equation}
    \ddot q-\frac{9}{4}H^2q+\lambda q^3=\lambda(1-e^{-3Ht})q^3.
\end{equation}
It is easy to see that, for $t\rightarrow\infty$, we get the limit equation
\begin{equation}
    \ddot q_{\infty}-\frac{9}{4}H^2q_{\infty}=0.
\end{equation}
This equation expresses the decay of the quintessence field as time runs to infinity. In the very early stages, the rhs becomes negligible small as the exponential is very near 1. Then, we can use perturbation theory. We will get
\begin{eqnarray}
\label{eq:pert}
   \ddot q_0-\frac{9}{4}H^2q_0+\lambda q_0^3&=&0, \nonumber \\
   \ddot q_1-\frac{9}{4}H^2q_1+3\lambda q_0^2q_1&=&\lambda(1-e^{-3Ht})q_0^3, \nonumber \\
   &\vdots&.
\end{eqnarray}
The leading order equation is a Higgs-like equation and admits a constant solution given by
\begin{equation}
    q_0(t)=\sqrt{\frac{9H^2}{4\lambda}}.
\end{equation}
Then, we have to solve
\begin{equation}
    \ddot q_1+\frac{9}{2}H^2q_1=\left(\frac{9H^2}{4\lambda^\frac{1}{3}}\right)^\frac{3}{2}(1-e^{-3Ht}).
\end{equation}
We get the solution
\begin{equation}
    q_1(t)=\frac{3H}{4\sqrt{\lambda}}-\frac{H}{4\sqrt{\lambda}}e^{-3Ht}+A\sin\left(\frac{3}{\sqrt{2}}Ht+\chi\right),
\end{equation}
being $A$ and $\chi$ two integration constants to be fixed by initial conditions. We emphasize that, in this approximation, $e^{-3Ht}\approx 1$. Finally, we get
\begin{equation}
    \phi(t)=\frac{9H}{4\sqrt{\lambda}}e^{-\frac{3}{2}Ht}-\frac{H}{4\sqrt{\lambda}}e^{-\frac{9}{2}Ht}+Ae^{-\frac{3}{2}Ht}\sin\left(\frac{3}{\sqrt{2}}Ht+\chi\right)+\ldots.
\end{equation}

\subsection{Non-constant solution}

We work again in the approximation having $e^{-3Ht}$ varying slowly in time in eq.(\ref{eq:q(t)}). We will get the solution \cite{Frasca:2009bc}
\begin{equation}
    q(t)\approx\pm\sqrt{\frac{3}{2\tilde\lambda(t)}}H\operatorname{dn}\left(m_0t+\chi,i\right)
\end{equation}
being dn a Jacobi elliptical function,
\begin{equation}
    \tilde\lambda(t)=\lambda e^{-3Ht},
\end{equation}
and
\begin{equation}
    m_0^2=\frac{3}{4}H^2.
\end{equation}
This will give finally
\begin{equation}
    \phi(t)=e^{-\frac{3}{2}Ht}q(t)=\pm\sqrt{\frac{3}{2\lambda}}H\operatorname{dn}\left(m_0t+\chi,i\right).
\end{equation}
Therefore, we recognize that this approximation is somewhat too strong. Such a solution represents oscillations around the constant value $\sqrt{\frac{3}{2\lambda}}H$ as the dn function is never zero. This constant is proportional to the value given in the preceding section for the constant solution.


A good asymptotic approximation for both the limit $Ht\gg 1$ to the solution of eq.(\ref{eq:q(t)}) is the following
\begin{equation}
    q(t)=\sqrt{\frac{3}{2\lambda}}H\operatorname{dn}\left(\frac{1}{2\sqrt{3}}(1-e^{-m_0t})+\chi,i\right).
\end{equation}
This will give
\begin{equation}
    \phi(t)=\sqrt{\frac{3}{2\lambda}}He^{-\frac{3}{2}Ht}\left[\operatorname{dn}\left(\frac{1}{2\sqrt{3}}(1-e^{-m_0t})+\chi,i\right)
    -\operatorname{dn}\left(\frac{1}{2\sqrt{3}}+\chi,i\right)\right].
\end{equation}
The opposite limit, $Ht\ll 1$, yields instead
\begin{equation}
    \phi(t)=\sqrt{\frac{3}{2\lambda}}He^{-\frac{3}{2}Ht}\operatorname{dn}\left(m_0t+\chi,i\right).
\end{equation}
This means that, in the long run, the solution tends to be nearly constant.


\medskip

\medskip

\section{Conclusions\label{conc}}
We showed that the dark energy or the late-time accelerated expansion of the universe that we observe can be accounted by the Higgs field of quintessence type in the non-perturbative regimes. We used a set of exact solutions of Higgs field theory obeying a massive dispersion relation for this purpose. We summarize below the salient features of our study: 
\begin{itemize}
    \item We showed Higgs with quartic potential with certain choices of the background solution via Jacobi elliptical function behave as Dark Energy. This depends on the choice of a single parameter, the phase $\theta$, that could have some value singled out at the very start of the birth of the Universe 
    (see Fig. 1).
    \item Here we introduced a novel iterative approach to non-local Higgs extensions and found the mass gap through a standard Fourier series. Also, through this simple technique, it was straightforwardly shown the consistency of our preceding derivation \cite{Frasca:2020jbe} that is distinctly different from the one previously studied making simpler the appearance of the mass gap in the theory.
    \item Similarly to the local case, where complete dark energy corresponds to the values of a single parameter, the phase $\theta=\pm(2n+1)K(i)$ with $n\in\mathbb{Z}$ (see Fig.~\ref{fig1}), we showed in non-local infinite-derivative ghost-free extensions of the local Higgs theory, that dark energy can be recovered again based on the choices of a single parameter, $\theta$, 
    (see Fig.~\ref{fig2}), for the values $\theta=\pm\pi/2+2n\pi$ being $n\in\mathbb{Z}$, assuming time variations on very large time scales. The behavior we recover is quite similar in both the local and non-local case.
    \item These ideas can be easily extended to other non-local theories like a Lee-Wick model that are particular cases of our most general formulation \cite{Frasca:2022duz} and dark energy will occur generally for such theories (see Eqn. (41)).
\end{itemize}
Summing up in this study we have presented an intriguing possibility,
that the background solution of the Higgs potential in standard field theory and higher-derivative field theories, involving Jacobi elliptical function leads to quintessence dark energy and that can yield the proper equation of state just depending on the values of a single parameter, the phase $\theta$. Further thorough and comprehensive analysis is required to verify if
this is truly viable, \textit{i.e.} if this leads to the currently observed distances and growth rates however that is  beyond the reach of the present manuscript and we leave it for future work.

\section*{Acknowledgement}
AG thanks Florian Nortier for comments. AK is supported by FCT Portugal investigator project IF/01607/2015.

\end{document}